# On graphs without a $C_4$ or a diamond

Elaine M. Eschen [*]    Chính T. Hoàng [†]    Jeremy P. Spinrad [‡]

R. Sritharan [§]

June 19, 2018

## Abstract

We consider the class of ($C_4$, diamond)-free graphs; graphs in this class do not contain a $C_4$ or a diamond as an induced subgraph. We provide an efficient recognition algorithm for this class. We count the number of maximal cliques in a ($C_4$, diamond)-free graph and the number of $n$-vertex, labeled ($C_4$, diamond)-free graphs. We also give an efficient algorithm for finding a largest clique in the more general class of (house, diamond)-free graphs.

[*]Elaine.Eschen@mail.wvu.edu, Lane Department of Computer Science and Electrical Engineering, P.O. Box 6109, West Virginia University, Morgantown, WV 26506.

[†]choang@wlu.ca, Department of Physics and Computing, Wilfrid Laurier University, Waterloo, Canada. Acknowledges support from NSERC of Canada.

[‡]spin@vuse.vanderbilt.edu, Department of Electrical Engineering and Computer Science, Vanderbilt University, Nashville, TN 37235.

[§]srithara@notes.udayton.edu, Computer Science Department, The University of Dayton, Dayton, OH 45469. Acknowledges support from the National Security Agency.

# 1 Introduction and motivation

In this paper we consider simple, connected, undirected graphs. A *diamond* is the graph obtained from $K_4$ by deleting an edge and $C_4$ is the cycle on four vertices. A graph is ($C_4$, diamond)-free if it contains neither a $C_4$ nor a diamond as an induced subgraph. In general, we use $\mathcal{F}$-*free* to refer to the class of graphs whose members do not contain any graph in the set $\mathcal{F}$ as an induced subgraph.

We use $n$ and $m$ to denote the number of vertices and edges, respectively, in a graph $G = (V, E)$. For a vertex $v$ in graph $G$, $N(v) = \{u \in V \mid uv \in E\}$ and $d(v) = |N(v)|$ denotes the degree of $v$ in $G$. We use $\Delta(G)$, or simply $\Delta$, to represent the maximum vertex degree in a graph $G$. When two vertices $u$ and $v$ are adjacent, we say $u$ *sees* $v$, or equivalently, $v$ *sees* $u$. For a subset $S$ of vertices of $G$, $G[S]$ denotes the subgraph induced by $S$ in $G$. $P_k$ is the path on $k$ vertices. A *house* is the complement of a $P_k$.

There are a number of important open questions regarding $C_4$-free graphs. Farber [4] showed that $C_4$-free graphs contain $O(n^2)$ maximal cliques, but it is not known whether this bound is tight. In general, it is not clear how to generalize results on chordal graphs to $C_4$-free graphs. Other open questions on $C_4$-free graphs include the total size of all maximal cliques, existence of a vertex that is contained in a small number of maximal cliques, and recognition algorithms.

Let $F$ be any graph on four vertices that is not $P_4$, $C_4$ or the complement of $C_4$. We have shown that recognizing $F$-free graphs is at least as hard as recognizing triangle-free graphs via simple $O(n^2)$-time reductions from recognizing triangle-free graphs. The best known algorithm for recognizing triangle-free graphs has time complexity $O(MM)$, where MM is the time required to multiply two $n$ by $n$ matrices; currently, the best known algorithm for dense graphs uses $O(n^{2.376})$ time [1, 6]. $P_4$-free graphs can be recognized in linear time [2]. A $C_4$-free graph recognition algorithm that beats matrix multiplication or a reduction from triangle-free graph recognition remains open.

This paper shows that we can get more precise results if we forbid the diamond as well as the $C_4$. Another way to characterize the class of ($C_4$, diamond)-free graphs is that a graph is in this class if and only if every nonadjacent pair of vertices has at most one common neighbor. A general technique we employ is to use a vertex degree threshold to balance work (or space) between the low and high degree vertices. If we have a low degree vertex, we can afford certain operations; alternately, if we have a high degree vertex, we can afford certain other operations. For ($C_4$, diamond)-free graphs, we give an exact bound on the number of maximal cliques. The techniques we employ are useful for solving other problems as well, such as improving the time for recognizing the class. Since the problems of computing the chromatic number and a largest independent set are NP-hard for graphs of large girth [3, 5], they remain so for the class of ($C_4$, diamond)-free graphs.

## 1.1 The projective plane graph

We provide a construction for a $C_4$-free bipartite graph on $2n$ vertices, referred to henceforth as the *projective plane graph*, that has $\Theta(n\sqrt{n})$ maximal cliques whose sizes sum to $\Theta(n\sqrt{n})$. This graph will be subsequently used in arguments for lower bounds.

Consider a finite projective plane $D$ of order $p$ with $p^2 + p + 1$ lines and $p^2 + p + 1$ points. We construct a bipartite graph $G$ based on $D$ as follows: vertices of the color class $X$ correspond to the lines of $D$ and the vertices of the color class $Y$ correspond to points of $D$. For vertex $x$ corresponding to line $L_x$ and vertex $y$ corresponding to point $P_y$, $x$ is adjacent to $y$ in $G$ if and only if $P_y$ lies on $L_x$. Since, $D$ is a finite projective plane, any two vertices of $G$ have at most one neighbor in common. Thus, $G$ is a $C_4$-free graph. Further, for any vertex $x \in X$, $d(x)$ is $p + 1$. Taking $n$ to be $p^2 + p + 1$, it is seen that $G$ has $\Theta(n\sqrt{n})$ maximal cliques whose sizes sum to $\Theta(n\sqrt{n})$.

## 2 Structure and the maximal cliques

**Lemma 2.1** *Suppose $v$ is a vertex of the diamond-free graph $G$. Then, $G[N(v)]$ is the disjoint union of cliques.*

**Proof of Lemma 2.1.** Since $G$ is diamond-free, $G[N(v)]$ does not contain a $P_3$ as an induced subgraph, and hence, is the disjoint union of cliques. □

**Lemma 2.2** *Suppose $v$ is a vertex of the diamond-free graph $G$. Then, $v$ is in at most $d(v)$ maximal cliques of $G$ and the sum of the sizes of these cliques is $O(d(v))$.*

**Proof of Lemma 2.2.** By Lemma 2.1, $G[N(v)]$ is the disjoint union of cliques. Therefore, $v$ is in at most $d(v)$ maximal cliques of $G$ and also the sum of the sizes of these maximal cliques is at most $2d(v)$. □

**Lemma 2.3** *Let $G$ be a ($C_4$, diamond)-free graph and $v$ be a vertex of $G$. Then, $N(v)$ induces a disjoint union of cliques, and for any $w \in N(v)$, $N(w) - N(v) - \{v\}$ induces a disjoint union of cliques. Furthermore, for a vertex $w \in N(v)$ and $x \in N(w) - N(v) - \{v\}$, the only neighbor that $v$ and $x$ have in common is $w$.*

**Proof of Lemma 2.3.** Let $v, w, x$ and $G$ be as in the statement of the lemma. The fact that $N(v)$ and $N(w) - N(v) - \{v\}$ each induce a disjoint union of cliques follows from Lemma 2.1. For vertices $w \in N(v)$ and $x \in N(w) - N(v) - \{v\}$, if a vertex $y \neq w$ were also a common neighbor of $v$ and $x$, then $\{v, x, y, w\}$ induces either a $C_4$ or a diamond in $G$. □

**Corollary 2.1** *For a ($C_4$, diamond)-free graph $G$ and vertex $v$ of $G$, let $A = \{wx \in E(G) \mid w \in N(v) \text{ and } x \notin N(v)\}$. Then, $|A| \leq n$.*

**Lemma 2.4** *Suppose $v$ is a vertex of the ($C_4$, diamond)-free graph $G$. The sum of the sizes of the maximal cliques of $G$ that contain members of $\{v\} \cup N(v)$ is $O(n)$. Moreover, all such maximal cliques can be enumerated in $O(m+n)$ time.*

**Proof of Lemma 2.4.** By Lemma 2.2, the sum of the sizes of the maximal cliques containing $v$ is $O(n)$. Next, we show that the sum of the sizes of maximal cliques of $G$ each of which does not contain $v$, but contains some neighbor of $v$, is also $O(n)$. By Lemma 2.3, a maximal clique of $G$ that does not contain $v$ cannot contain two neighbors of $v$; also, by Lemma 2.3, among the maximal cliques of $G$ that do not contain $v$, a maximal clique containing $x \in N(v)$ and a maximal clique containing $y \in N(v)$ cannot have any vertex in common. It follows that for a vertex $w \in N(v)$, the maximal cliques of $G$ that contain $w$, but do not contain $v$, are precisely the disjoint cliques induced by $N(w) - N(v) - \{v\}$ with $w$ added to each of them. Therefore, applying Lemma 2.3 and Corollary 2.1, we can conclude that the total size of the maximal cliques of $G$ that do not contain $v$, but contain some neighbor of $v$, is also $O(n)$. A straightforward algorithm, based on Lemma 2.3, can be used to enumerate all the relevant maximal cliques in $O(m+n)$ time. □

We note that the statement of Lemma 2.4 is not true for the class of diamond-free graphs (consider $K_{n,n}$) or for the class of $C_4$-free graphs (consider the graph obtained from the projective plane graph by adding a universal vertex).

**Theorem 2.1** *Let $G$ be a ($C_4$, diamond)-free graph on $n$ vertices. The sum of the sizes of the maximal cliques of $G$ is $O(n\sqrt{n})$; hence, the number of maximal cliques of $G$ is also $O(n\sqrt{n})$. Further, there exist ($C_4$, diamond)-free graphs on $n$ vertices containing $\Omega(n\sqrt{n})$ maximal cliques; hence, the sum of their sizes is also $\Omega(n\sqrt{n})$.*

**Proof of Theorem 2.1.** Let $S$ be the sum of the sizes of maximal cliques of $G$. To bound $S$, we consider classes of vertices, assess the total size of the maximal cliques containing vertices in the class, and then eliminate these vertices from future consideration. When we consider a vertex $v$ of degree at most $\sqrt{n}$, since by Lemma 2.2 the total size of the maximal cliques containing $v$ is $O(\sqrt{n})$, we assess $O(\sqrt{n})$ to $S$ and remove $v$ from consideration. Clearly, the total assessment to $S$ due to all such vertices $v$ is $O(n\sqrt{n})$. When we consider a vertex $v$ with degree more than $\sqrt{n}$, since by Lemma 2.4 the total size of the maximal cliques containing a member of $\{v\} \cup N(v)$ is $O(n)$, we assess $O(n)$ to $S$ and remove $v$ as well as all members of $N(v)$ from consideration. Since at least $\sqrt{n} + 1$ vertices are thus removed from consideration, the number of times we can assess $S$ with $O(n)$ in this manner is $O(\sqrt{n})$; therefore, the total of all such assessments is also $O(n\sqrt{n})$. Finally, the projective plane graph demonstrates that the bound is tight. □

## 3 The recognition problem

Note that graph $G$ is ($C_4$, diamond)-free if and only if for every pair $\{x, y\}$ of nonadjacent vertices of $G$, $|N(x) \cap N(y)| \leq 1$. Therefore, whether a given graph is ($C_4$, diamond)-free can be tested using matrix multiplication in $O(n^{2.376})$ time [1, 6]. In this section, we present an $O(m^{\frac{2}{3}}n)$-time algorithm to recognize ($C_4$, diamond)-free graphs. We first show that ($C_4$, diamond)-free graphs can be recognized in $O(m\Delta)$ time. This will be used as a subroutine in our $O(m^{\frac{2}{3}}n)$-time algorithm.

### 3.1 An $O(m\Delta)$-time algorithm

Note that $v$ is a vertex of degree two in a diamond or a $C_4$ if and only if some non-neighbor of $v$ shares at least two neighbors with $v$. Therefore, in order to check whether a particular vertex $v$ is part of a diamond or a $C_4$, we perform a breadth-fist search starting from $v$ until all the vertices at distance two from $v$ are marked. Also, during the search the moment we discover that a vertex at distance two from $v$ is adjacent to two vertices that are at distance one from $v$, we stop since $v$ is part of a $C_4$ or a diamond. The overall cost of running such a search from each vertex is $O(n + \Sigma d(v) + \Delta \Sigma d(v)) = O(m\Delta)$, where the sums are taken over $V(G)$.

## 3.2 An $O(m^{\frac{2}{3}}n)$-time algorithm

Our basic strategy is that when the maximum degree of the graph is "high", we pick a vertex $v$ with such a degree and eliminate $v$ as well as all its neighbors from consideration at a cost of $O(m)$ time. When the maximum degree of the remaining graph eventually becomes "not high", we can afford to run the $O(m\Delta)$-time algorithm on it. The suitable threshold $f$ for the maximum degree will be determined later so as to optimize the cost of the algorithm.

**Algorithm** *Recognition*
**Input**: Graph $G$
**Output**: *yes* if $G$ is ($C_4$, diamond)-free and *no* otherwise
(1)   **while** $\Delta(G) > f$ **do**
        Let $v$ be a vertex of degree $\Delta(G)$.

        /* Eliminate $v$ and $N(v)$ from consideration */
        Perform a breadth-first search from $v$ until all vertices at
        distance 3 from $v$ are marked.
        Let $T$ be the resulting tree with 4 levels.
        Let $L_i = \{w \mid \text{distance between } v \text{ and } w \text{ is } i\}$.
        As the search progresses do the following:

(1a)     **if** $L_1$ does not induce a disjoint union of cliques **then**
           return (no)
        **endif**
(1b)     **if** a vertex in $L_2$ sees two vertices in $L_1$ **then**
           return (no)
        **endif**
(1c)     **for** every vertex $w \in L_1$ **do**
           **if** $N(w) \cap L_2$ does not induce a disjoint union of cliques **then**
               return (no)
           **endif**
        **endfor**
(1d)     **for** every edge $xy$ with $x \in L_2$ and $y \in L_2$ **do**
           Let $x'$ be the only neighbor of $x$ in $L_1$.

    Let $y'$ be the only neighbor of $y$ in $L_1$.
    **if** $x'$ sees $y'$ **then**
     return (no)
    **endif**
   **endfor**
(1e)   **if** $x \in L_2$ sees $a \in L_2$ and $b \in L_2$ such that $\text{parent}(a) = \text{parent}(b)$, but $\text{parent}(x) \neq \text{parent}(a)$ **then**
   return (no)
  **endif**
(1f)   **if** a vertex in $L_3$ sees two vertices in $L_2$ **then**
   return (no)
  **endif**
  Delete $N(v) \cup \{v\}$ from $G$ and update the degrees of the remaining vertices.
 **endwhile**
(2)  Run the $O(m\Delta)$ time algorithm on remaining graph.

**Theorem 3.1** *Algorithm Recognition is correct.*

**Proof of Theorem 3.1.** In every case (in both Step (1) and Step (2)) that the algorithm returns *no*, the input graph $G$ contains an induced $C_4$ or an induced diamond. It remains to argue that if the algorithm returns *yes*, $G$ is indeed ($C_4$, diamond)-free.

*Claim*: If a vertex $v$ (the root of the breadth-first search) and its neighborhood $N(v)$ are removed from $G$ in Step (1), then no vertex in $v \cup N(v)$ is part of an induced $C_4$ or an induced diamond in $G$.

*Proof*: Vertex $v$ is of degree three in a diamond if and only if $N(v)$ induces a $P_3$. Therefore, Step (1a) is sufficient to verify whether $v$ is a degree three vertex in a diamond. Vertex $v$ is of degree two in a diamond or a $C_4$ if and only if some non-neighbor of $v$ shares at least two common neighbors with $v$. Vertex $v$ can have a common neighbor with a non-neighbor only if the non-neighbor is in $L_2$. Therefore, Step (1b) is sufficient to verify whether $v$ is a degree two vertex in a diamond or a $C_4$.

Let $w$ be a vertex in $N(v)$. Note that after Step (1b), it is established for $G$ that (i) $L_1$ induces a disjoint union of cliques and (ii) no vertex in $L_2$ sees two vertices in $L_1$. It follows from (i) that if $w$ is of degree three in a diamond, then the diamond must include a neighbor of $w$ that is in $L_2$. Then all the neighbors of $w$ in the diamond must be in $L_2$, since no vertex in $L_2$ can see both $w$ and another vertex in $L_1$. Therefore, Step (1c) is sufficient to verify whether $w$ is a degree three vertex in a diamond.

Vertex $w$ is of degree two in a diamond or a $C_4$ if and only if some non-neighbor $u$ of $w$ shares at least two common neighbors with $w$. We consider the possible positions for $w$, $u$, and their common neighbors. Vertex $w$ can have a common neighbor with a non-neighbor only if the non-neighbor is in $L_1, L_2$, or $L_3$. If $u$ is in $L_1$ and has a common neighbor with $w$ in $L_1$, then (i) is violated. If $u$ is in $L_1$ and has a common neighbor with $w$ in $L_2$, then (ii) is violated. So suppose $u$ is in $L_2$. By (ii), $u$ has only one neighbor in $L_1$. The case that $w$ and $u$ have a common neighbor in $L_1$ and a common neighbor in $L_2$ is covered by Step (1d) of the algorithm. The case that $w$ and $u$ have two common neighbors in $L_2$ is covered by Step (1e). Finally, suppose $u$ is in $L_3$. Vertices $w$ and $u$ can have common neighbors only in $L_2$. Therefore, Step (1f) is sufficient for this case. □

By the claim, the algorithm "safely" removes vertices from the input graph until the maximum degree is below the threshold. Then the algorithm of Section 3.1 is run. The correctness of this algorithm is given in Section 3.1. □

**Theorem 3.2** *Algorithm Recognition can be implemented to run in $O(m^{\frac{2}{3}}n)$ time.*

**Proof of Theorem 3.2.** First, we show that Step (1) of the algorithm can be done in $O(m)$ time. Step (1a) is done by checking if every component of the subgraph induced by $L_1$ is a clique. We also mark the vertices in $L_1$ with the label of their component. Step (1b) can be done by marking vertices appropriately during the search. If the algorithm progresses to Step (1c), then no vertex in $L_2$ sees two vertices in $L_1$. Therefore, for any two distinct vertices in $L_1$, the sets of their children in $T$ are disjoint. Hence, Step (1c) can be handled in a manner similar to Step (1a). For Step (1d), whether or not $x'$ sees $y'$ can be decided in constant time by checking if they belong to the same component of subgraph induced by $L_1$. Step (1e) can be done by marking vertices appropriately during the search. Finally, Step (1f) is similar to Step (1b). Therefore, Step (1) of the algorithm can be done in $O(m)$ time. Since at least $f + 1$ vertices are discarded from the graph when Step (1) is done, the number of times Step (1) is done is at most $\frac{n}{f}$. Therefore, the total time spent by the algorithm in Step (1) is $O(m\frac{n}{f})$. Let $G'$ be the graph remaining after Step (1). Then, $\Delta(G') \leq f$, and hence, the number of edges in $G'$ is $O(nf)$. Therefore, Step (2) of the algorithm runs in $O(nf^2)$ time. Balancing the total cost of Step (1) with the cost of Step (2), we arrive at $f = \Theta(m^{\frac{1}{3}})$. It follows that the entire algorithm runs in $O(m^{\frac{2}{3}}n)$ time. □

# 4 The maximum clique problem on (house, diamond)-free graphs

In this section, we present an $\mathrm{O}(m^{\frac{2}{3}}n)$-time algorithm to compute a largest clique in a given (house, diamond)-free graph. We basically enumerate cliques, including all maximal cliques, of the given graph to compute the largest one. We use ideas similar to that of the recognition algorithm of Section 3 to control the complexity.

**Definition 4.1** *Let $v$ be a vertex of a graph $G$. We denote by $N_i(v)$ the set of vertices of distance $i$ from $v$.*

**Lemma 4.1** *Let $G$ be a (house, diamond)-free graph and let $y$, $v$ be two vertices at distance two in $G$ that have at least two common neighbors. Then for any common neighbor $x$ of $y$ and $v$, $xy$ and $xv$ are maximal cliques of $G$.*

**Proof of Lemma 4.1.** Let $G$, $y$, $v$ be defined as in the lemma. Let $Q$ be the set of common neighbors of $v$ and $y$. For any two vertices $x_1$, $x_2$ of $Q$, $x_1$ does not see $x_2$ since $G$ is diamond-free. Consider a vertex $x$ of $Q$ and suppose that $xy$ is not a maximal clique of $G$; i.e., there is a vertex $u \in N(v) \cup N_2(v)$ that sees both $x$ and $y$. We have $u \notin N(v)$ since $G$ is diamond-free. Thus, $u$ is in $N_2(v)$. Consider a vertex $x'$ different from $x$ of $Q$. Vertex $u$ does not see $x'$, for otherwise $\{x, u, y, x'\}$ induces a diamond. But, now $\{v, x, u, y, x'\}$ induces a house. Thus, a contradiction is reached.□

**Theorem 4.1** *For a vertex $v$ of a (house, diamond)-free graph, there is a linear time algorithm to list all maximal cliques that intersect $N(v) \cup \{v\}$.*

**Proof of Theorem 4.1.** Consider a vertex $v$ of a (house, diamond)-free graph $G$. Let $\mathcal{C}$ be the set maximal cliques of $G$ that intersect $N(v) \cup \{v\}$. We describe a linear time algorithm to list all members of $\mathcal{C}$.

**Algorithm** *Enumerate-maximal-cliques*
**Input**: (house, diamond)-free graph $G$ and vertex $v$
**Output**: The set of maximal cliques of $G$ that intersect $N(v) \cup \{v\}$
(1) Compute $N(v)$ and $N_2(v)$.
(2) Compute the maximal cliques $K_1, \ldots, K_r$ of $G[N(v)]$.
(3) Output the maximal cliques $K_i \cup \{v\}$.

(4) **for** each clique $K_i$ **do**
      **for** each vertex $x \in K_i$ **do**
(4a)       Compute the set $A_x = N(x) \cap N_2(v)$.
(4b)       Compute the maximal cliques $R_1, \ldots, R_s$ of $G[A_x]$.
(4c)       Output the maximal cliques $R_i \cup \{x\}$.
      **endfor**
    **endfor**

If there is a clique $K \in \mathcal{C}$ that is not contained in $N(v) \cup \{v\}$, then $K$ must contain exactly one vertex $x$ in $N(v)$ since $G$ is diamond-free. Thus, $K - \{x\}$ will be recognized as a clique $R_i$ in Step (4b). So, the algorithm is correct.

Clearly, Steps (1) and (3) can be implemented to run in linear time. Step (2) can be implemented to run in linear time due to Lemma 2.1; a straightforward search algorithm can be used to enumerate the disjoint cliques induced by $G[N(v)]$. Next we show that Step (4) can be implemented to run in linear time. We we show that the adjacency list of each vertex in $N(v) \cup N_2(v)$ need only be scanned at most once. This is obviously true for vertices in $N(v)$. We will implement Step (4b) as follows. We pre-compute in linear time the set $Y \subseteq N_2(v)$ such that each vertex in $Y$ has at least two neighbors in $N(v)$. Consider a vertex $y \in A_x$. In the case that $y \notin Y$, due to Lemma 2.3, we can scan the adjacency list of $y$ to find a clique $R_i$. Note this will not be done again for any other vertex $x' \in N(v)$. In the case that $y \in Y$, by Lemma 4.1, $y$ has no neighbor in $A_x$. Thus, $y$ is a singleton maximal clique of $G[A_x]$, and its adjacency list does not need to be scanned. This proves the algorithm runs in linear time. □

We can now describe our algorithm to find a largest clique in a (house, diamond)-free graph.

**Algorithm** *Maxclique*
**Input**: (house, diamond)-free graph $G$
**Output**: A largest clique in $G$
(1) **while** $\Delta(G) > f$ **do**
    Let $v$ be a vertex of degree $\Delta(G)$.
    Enumerate all the maximal cliques of the remaining graph that contain members of $N(v) \cup \{v\}$, noting the largest clique.
    Delete $N(v) \cup \{v\}$ and update the degrees of the remaining vertices.
  **endwhile**
(2) Let $G'$ be the remaining graph.
    **for** every vertex $v$ in $G'$ **do**

       Enumerate all maximal cliques of $G'$ containing $v$, noting the largest clique.

   **endfor**

(3)  Output the largest clique found.

**Theorem 4.2** *Algorithm Maxclique is correct and it runs in $O(m^{\frac{2}{3}}n)$ time.*

**Proof of Theorem 4.2.** It is easily seen that the algorithm will generate every maximal clique of the (house, diamond)-free graph $G$ given as input.

By Theorem 4.1, a single invocation of Step (1) of the algorithm can be done in $O(m)$ time. Since at least $f + 1$ vertices are discarded every time Step (1) is executed, the overall time spent on Step (1) is $O(m\frac{n}{f})$. For a vertex $v$ of $G'$, a straightforward search algorithm can be used to enumerate the disjoint cliques induced by $N(v)$ in $G'$, due to Lemma 2.1. Since the sum of the degrees of the neighbors of $v$ in $G'$ is $O(f^2)$, this takes $O(f^2)$ time. Therefore, Step (2) of the algorithm can be done in $O(nf^2)$ time. The rest of the analysis is identical to that in the proof of Theorem 3.2.   □

## 5   Number of labeled ($C_4$, diamond)-free graphs

In this section, we show that the number of labeled ($C_4$, diamond)-free graphs on $n$ vertices is $2^{O(n\sqrt{n}\log n)}$.

**Theorem 5.1** *The number of labeled ($C_4$, diamond)-free graphs on $n$ vertices is $2^{O(n\sqrt{n}\log n)}$.*

**Proof of Theorem 5.1.** We will map each labeled ($C_4$, diamond)-free graph on $n$ vertices to a distinct binary representation (from which the edge set can be determined) that uses $O(f(n))$ bits. From this we can conclude there are $2^{O(f(n))}$ labeled ($C_4$, diamond)-free graphs on $n$ vertices. We do this by encoding the adjacencies of a ($C_4$, diamond)-free graph $G$ on $n$ vertices by a set of $n$ lists (one list per vertex) where the length of each list is $O(\sqrt{n})$. Some of the members of a list are neighbors of the vertex, while the rest are non-neighbors of the vertex. We also store the vertex sets of $O(\sqrt{n})$ subgraphs of $G$ in order to resolve neighbors when non-neighbors are stored.

Let $G$ be a ($C_4$, diamond)-free graph on $n$ vertices. Consider each vertex of $G$. If a vertex $v$ has degree at most $\sqrt{n}$, then we add all the neighbors of $v$ to $v$'s list and remove $v$ from consideration. Otherwise, if $v$ has at most $\sqrt{n}$ non-neighbors, then we add all non-neighbors of $v$ to $v$'s list and remove $v$ from consideration. If the non-neighbors of any vertex are stored, then we also store the vertex set of $G$.

Now all remaining vertices have more than $\sqrt{n}$ neighbors and more than $\sqrt{n}$ non-neighbors. We choose an arbitrary vertex $v$ and decompose $G$ into $G_1$, the subgraph of $G$ induced by $N(v) \cup \{v\}$, and $G_2$, the subgraph of $G$ induced by $V(G) - N(v) - \{v\}$. Note that $G_1$ and $G_2$ each have more than $\sqrt{n}$ vertices and the vertex sets of $G_1$ and $G_2$ are disjoint. By Lemma 2.3, a vertex in $G_2$ has at most one neighbor in $G_1$. For each such edge $xy$ with $x$ in $G_1$ and $y$ in $G_2$, we add $x$ to $y$'s list.

We then process adjacencies within $G_1$ and $G_2$ recursively using $\sqrt{n}$ (where $n$ is the number of vertices in $G$) as the degree bound, but computing neighborhoods in $G_i$. The decomposition can be naturally represented by a binary tree $T$ whose root corresponds to $G$, while the two subtrees correspond to $G_1$ and $G_2$. Note that every time a graph is split thus, each of $G_1$ and $G_2$ have more than $\sqrt{n}$ vertices. Therefore, the number of leaves of $T$ is at most $\sqrt{n}$. Hence, the number of decomposition steps is also at most $\sqrt{n}$.

A vertex is removed from consideration exactly once and at this time at most $\sqrt{n}$ vertices are added to it's list. The list for a vertex increases in size by at most one when a decomposition step is done. Hence, the number of vertices added to the list of a particular vertex over all the splits is at most $\sqrt{n}$. Therefore, the list for every vertex has at most $2\sqrt{n}$ vertices. Since there are at most $\sqrt{n}$ decomposition steps, the sum of the sizes of the stored vertex sets for subgraphs of $G$ is $O(n\sqrt{n})$. Since the label of a single vertex can be encoded with $O(\log n)$ bits, the entire graph can be encoded with $O(n\sqrt{n} \log n)$ bits. Therefore, the number of labeled ($C_4$, diamond)-free graphs on $n$ vertices is $2^{O(n\sqrt{n}\log n)}$. □